\documentclass[aps,prx,reprint,twocolumn,floatfix]{revtex4-2}
\usepackage{iceberg}
\usepackage{amsmath}
\usepackage{amssymb}
\usepackage{amsthm}
\usepackage[section]{placeins}

\begin{document}

\title{The Pinnacle Architecture with fixed connectivity of degree eight}
\author{Paul Webster}
\affiliation{Iceberg Quantum}
\author{Tom Peham}
\affiliation{Iceberg Quantum}
\author{Lawrence Z.\ Cohen}
\affiliation{Iceberg Quantum}
\author{{\scriptsize\texttt{\{paul,tom.p,larry\}@iceberg-quantum.com}}}

\begin{abstract}
We show how the Pinnacle architecture can be adapted to be compatible with superconducting qubits, where the connectivity between qubits must be fixed at fabrication.
Specifically, we present a modified instantiation of the architecture that has fixed connectivity degree of eight.
With this instantiation, we show that a 2048-bit RSA integer can be factored in one month with approximately 120\,000 physical qubits, given a physical error rate of $10^{-3}$, code cycle time of 1 \textmu s and reaction time of 10 \textmu s.
\end{abstract}

\maketitle

\section{Introduction}
In Ref.~\cite{webster_pinnacle_2026}, it was shown that 2048-bit integers could be factored with fewer than 100\,000 physical qubits using the Pinnacle Architecture.
However, this result was based on a hardware-agnostic analysis, which left open the question of how adapting the architecture to specific hardware platforms may affect the estimated overhead.
In this work, we address this question with regards to superconducting qubits.

Superconducting qubits support sufficiently fast gate and measurement times to make code cycle times on the order of one microsecond plausible \cite{acharya_quantum_2025}.
However, unlike alternatives such as neutral atoms \cite{bluvstein_quantum_2022}, trapped ions \cite{pino_demonstration_2021} or spin qubits \cite{yoneda_coherent_2021}, they do not support flexible reconfiguration of qubits. 
Instead, the connectivity of qubits must be fixed at fabrication.
This necessitates that the fixed connectivity degree -- the maximum size of the set of qubits to which any qubit must be connected across the full operation of the architecture -- is small.
This is a property that is not satisfied by the instantiation of the Pinnacle Architecture presented in Ref.~\cite{webster_pinnacle_2026}, where the set of qubits to which any qubit must be connected at any given time is small, but this set changes over time as gadget positions are reconfigured.

We here address this issue by presenting a modified instantiation of the Pinnacle Architecture, which we refer to as the \textit{Frozen Pinnacle Architecture}. 
The Frozen Pinnacle Architecture has a fixed connectivity of degree eight.
This is comparable to the fixed connectivity degree of seven of the primary alternative quantum LDPC architecture -- the bicycle architecture of Ref.~\cite{yoder_tour_2025} (an instantiation of the extractor architecture of Ref.~\cite{he_extractors_2025}).
We show that the Frozen Pinnacle Architecture can support 2048-bit integer factoring in one month with approximately 120\,000 physical qubits.
This supports the conclusion that RSA-2048 factoring with devices of close to 100\,000 physical qubits in a feasible runtime is plausible using the Pinnacle Architecture.

In \cref{sec:background}, we review the relevant background.
In \cref{sec:architecture} we explain how each module of the Pinnacle architecture as it is instantiated in Ref.~\cite{webster_pinnacle_2026} can be modified to achieve a fixed connectivity degree of eight.
In \cref{sec:simulations}, we present numerical simulation results demonstrating that the Frozen Pinnacle Architecture achieves sufficiently low logical error rates for RSA-2048 factoring
In \cref{sec:resource-estimate}, we provide a resource estimate for RSA-2048 factoring on this architecture.

\section{Background}
\label{sec:background}
We take as our starting point the instantiation of the Pinnacle Architecture that supports factoring with a physical error rate of $p=10^{-3}$, a code cycle time of one microsecond and a reaction time of ten microseconds in less than one month using fewer than one hundred thousand physical qubits, as presented in Ref.~\cite{webster_pinnacle_2026}.
We refer to this throughout as the baseline architecture.
This follows the compilation of Ref.~\cite{gidney_how_2025} by using one working register, which is realised by a single processing unit which uses the gadget construction of Ref.~\cite{webster_explicit_2025} to allow any logical Pauli product measurement on the logical qubits of the working register to be implemented in a single logical cycle consisting of $O(d)$ rounds of syndrome extraction.
This is complemented by a magic engine that provides one $|\bar{T}\rangle$ state of infidelity $\leq 5\times 10^{-11}$ per logical cycle, to allow for universal quantum computing by Pauli-based computation \cite{bravyi_trading_2016,litinski_game_2019}.
A memory is also used to store the logical qubits on the input register; this consists of a set of code blocks along with a port which allows access to one memory window containing half of the logical qubits of a code block at a time.

Throughout the architecture, we use the generalised bicycle (GB) code family presented in Ref.~\cite{webster_explicit_2025,webster_pinnacle_2026}.
An $[[n,k,d]]$ instance of this family can be used to construct a code block of $n_{cb}=2n$ physical qubits (counting both data and ancilla qubits) with fixed connectivity degree six \cite{webster_pinnacle_2026}.
We focus on the $[[510,16,24]]$ code with lift $l=255$ and sets $A=\{0,39,55\}$ and $B=\{0,70,127\}$,
since this achieves a sufficiently low logical error rate for factoring with a physical error rate of $p=10^{-3}$ \cite{webster_pinnacle_2026}.
However, analogous constructions to those we present here can be constructed for other codes in the family.

Throughout, all gadgets we refer to are those from the gadget construction of Ref.~\cite{webster_explicit_2025}, which depends on the following structure.
The physical qubits of the code can be partitioned into two physical sectors of $l=255$ qubits each.
The code block supports $l$ cyclic shift automorphisms, corresponding to cyclic shifts of both physical sectors by $s\in\mathbb{Z}_l$ positions.
The logical qubits of the code can be partitioned into two equal logical sectors.
For each logical sector and each of $X$-type and $Z$-type operators, a seed operator can be chosen such that all $2^{k/2}-1$ logical operators of that type have representatives that are in the orbit of the seed operator under cyclic shift automorphisms \cite{webster_explicit_2025}.
This means that a gadget with fixed internal structure can measure any of the $2^{k/2}-1$ logical operators of one type on one logical sector depending on its position relative to the code block.
Moreover, such gadgets can be designed to have connectivity degree of no more than eight, and such that this remains true even when the gadget is connected to a code block and bridged to up to two other gadgets.

Each cyclic shift automorphism can be performed using existing connectivity of the code block.
Specifically, two layers of swaps between connected data and ancilla qubits allow a set of primitive cyclic shifts of each of the physical sectors by $a_i-a_j$ (mod $l$) or $b_i-b_j$ (mod $l$) for $i,j\in\{0,1,2\}$  \cite{bravyi_highthreshold_2024,yoder_tour_2025}.
By enumerating all cases, we find that every cyclic shift automorphism can be implemented by a sequence of no more than four such primitive shifts on the $[[510,16,24]]$ GB code used in this architecture.

\section{Architecture}
\label{sec:architecture}
\begin{figure}[t]
  \centering
  \includegraphics[width=\linewidth]{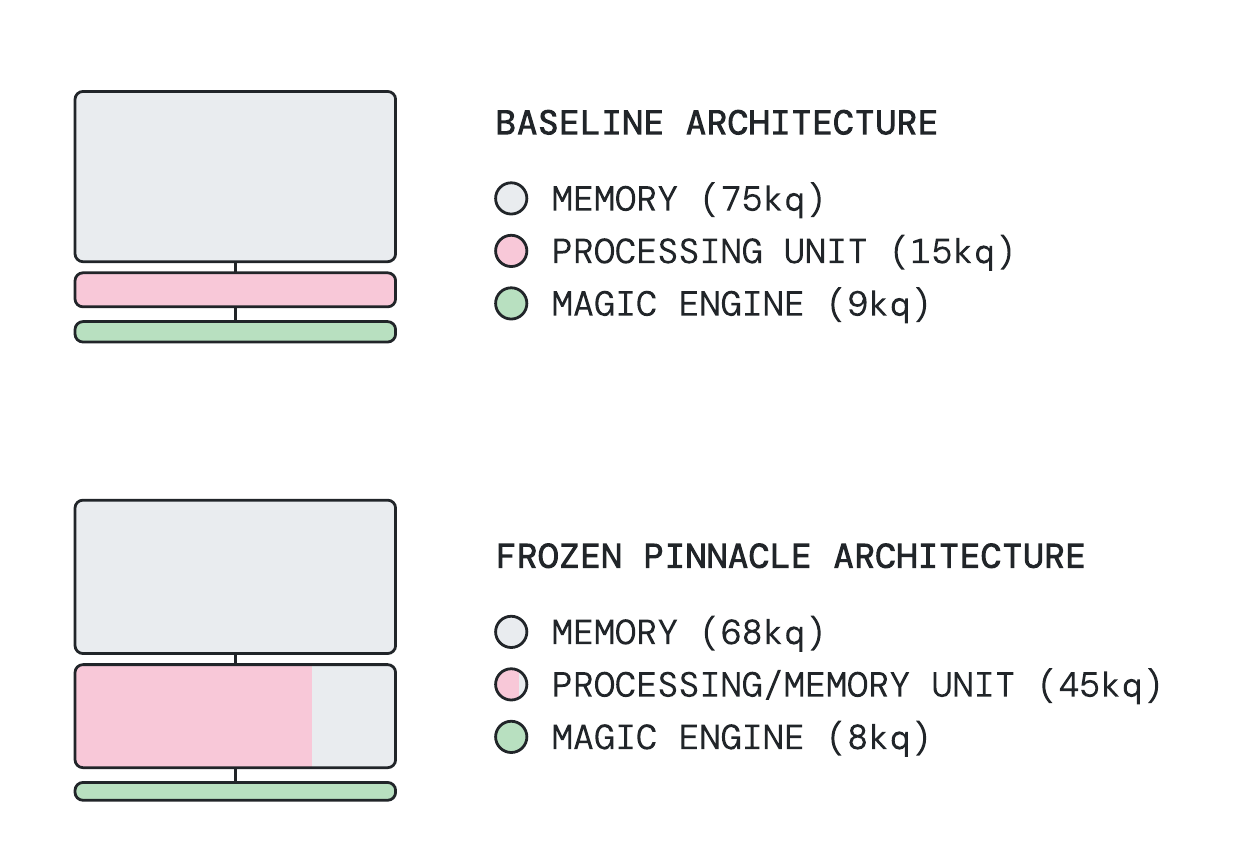}
  \caption{Comparison of the baseline architecture of Ref.~\cite{webster_pinnacle_2026}, which allows for RSA-2048 factoring with approximately 100\,000 qubits in one month and the Frozen Pinnacle Architecture presented here, which allows factoring with approximately 120\,000 qubits in the same time while ensuring a fixed connectivity degree of eight.
  The Processing/Memory Unit refers to a dual-use module made up of blocks with both a processing sector and a memory sector, as explained in \cref{sec:processing-unit} and \cref{sec:memory-sectors}.}
  \label{fig:architecture}
\end{figure}

In this section, we detail the Frozen Pinnacle Architecture, by considering in turn how each of the modules of the baseline architecture are modified for its construction.
\Cref{fig:architecture} shows a high-level representation of this architecture compared with the baseline architecture of Ref.~\cite{webster_pinnacle_2026}.

\subsection{Processing unit}
\label{sec:processing-unit}
A processing unit is constructed from bridging together $\beta\in\mathbb{N}$ processing blocks such that any logical Pauli product operator can be measured in a single logical cycle.

In the baseline architecture, processing blocks are constructed using four bridged gadgets, designed such that they measure the code's four seed operators.
Cyclically shifting these four seed operators relative to the code block then allows for arbitrary logical Pauli product measurements.
This is incompatible with the constraints of this paper, since it means that different logical measurements can result in gadget qubits being connected to any of the $n=510$ qubits of the code block. 

To achieve compatibility, we modify the construction as follows.
First, only the first logical sector of each code block is used for the processing block; we call this the \textit{processing sector}.
This means that each processing block provides only $k/2$ logical qubits instead of $k$.
However, the overhead impact of this can be mitigated by repurposing the second logical sector for memory (as explained in \cref{sec:memory-sectors}).
Second, $k/2$ gadgets are used to measure logical $\bar{Z}$-type operators on the processing sector, instead of only one gadget.
Each of these is connected at a fixed location such that each measures one element of a generating set, $\mathcal{Z}$, of the space of logical $\bar{Z}$-type operators on the processing sector.
Only a single $X$-type gadget is used, which is connected such that it measures the logical operator, $\bar{X}_s$, corresponding to the seed $X$-type operator.
In all logical cycles, the internal circuit for all of the $k/2+1$ gadgets is implemented, along with all required bridges between adjacent gadgets, ensuring that fixed connectivity across the gadget system is maintained.
However, for each gadget it can be independently chosen whether to activate the connections to the code block to measure the associated logical operator, or not to activate these connections in which case the gadget measures the logical identity.

An arbitrary logical Pauli operator, $\bar{P}$, on the processing unit is then measured in a single logical cycle as follows.
Express the restriction of $\bar{P}$ to the $i$th processing block as $\bar{P}_i\propto \tilde{X}_i\tilde{Z}_i$, where $\tilde{X}_i$ is an arbitrary logical $X$-type operator on the processing sector or the identity, and $\tilde{Z}_i$ is an arbitrary logical $Z$-type operator on the processing sector or the identity.
For each processing block, perform a cyclic shift automorphism, $U_{\tilde{X}_i}$.
If $\tilde{X}_i\neq \bar{I}$, this is chosen such that $U_{\tilde{X}_i}\bar{X}_sU_{\tilde{X}_i}^\dag=\tilde{X}_i$, which is possible because every logical $X$-type operator on the processing sector has a representative that is some cyclic shift of the seed operator; following this cyclic shift, activating the $X$-type gadget measures $\tilde{X}_i$.
If $\tilde{X}_i=\bar{I}$ then $U_{\tilde{X}_i}$ is chosen to be the identity.
Since $U_{\tilde{X}_i}$ is an automorphism that acts independently on the two logical sectors, then the fact that $\mathcal{Z}$ is a generating set for all logical $\bar{Z}$-type operators on the processing sector implies that $U_{\tilde{X}_i}\mathcal{Z}U_{\tilde{X}_i}^\dag$ is also a generating set for all logical $\bar{Z}$-type operators of the processing sector.
This means that, for any logical $\bar{Z}$-type operator, $\tilde{Z}_i$, there exists a subset of these gadgets that can be activated to measure $\tilde{Z}_i$.
We can therefore measure $\bar{P}_i$ by activating this subset of $Z$-type gadgets, along with the $X$-type gadget if $\tilde{X}_i\neq\bar{I}$.
The full logical operator $\bar{P}$ can therefore be measured by bridging the gadget systems of neighbouring processing blocks across the processing unit.

This construction can be realised with a fixed connectivity degree of eight.
Indeed, each gadget is connected to a code block in fixed position and, by bridging the gadgets in a linear chain within and between processing blocks, we can ensure that each gadget is bridged to at most two other gadgets.
As noted in \cref{sec:background}, this means we can ensure the gadgets all have fixed connectivity degree no greater than eight.
For the processing blocks, since the code block has fixed connectivity degree six, it suffices to ensure that no code block qubit (including ancilla qubits) is connected to more than two gadgets, which can be achieved with an appropriate basis and choice of operators.

The described operation of the processing unit requires a cyclic shift automorphism to be applied to each processing block at the start of each logical cycle; this is chosen to simultaneously undo the shift of the previous logical cycle and implement the required shift for the next logical cycle.
This marginally increases the time required per logical measurement, and hence the logical cycle time of the architecture.
Specifically, each such automorphism requires a sequence of at most four primitive automorphisms.
We allow for one round of syndrome extraction between each shift, and also account for the total time required for the gates of the shift itself by adding one further code cycle.
This increases the number of code cycles per logical cycle from $d_t$ in the baseline architecture to $d_t+4$.

\subsection{Memory}
\label{sec:memory}
\subsubsection{Memory Blocks}
As in the baseline architecture, we allow for dedicated memory blocks.
Each of these is a code block with two memory windows (each with $k/2$ logical qubits) corresponding to its two logical sectors.

We choose one memory block to be a \textit{port block}.
On this block, we assign two gadgets -- one measuring a physical $Z$-type operator on the first logical sector and the other measuring a physical $X$-type operator on the second logical sector.
We connect these gadgets to disjoint operators, which ensures that each physical qubit is in the support of at most one of them.
We define the logical basis of the second logical sector to be the dual of the usual choice.
This means that any $Z$-type logical operator on either logical sector can be measured by performing a cyclic shift and then activating one of the gadgets.
We bridge the first gadget to the second and the second gadget to the last gadget of a processing block at the end of the processing unit.
This suffices to allow the processing unit access to either memory window of the port block, while ensuring that each gadget is bridged to no more than two other gadgets.

To allow access to any window of a memory block, we add functionality for cyclic shifts of the blocks such that the data of any memory block can be moved into the port block.
To do this, we choose a perfect matching of the Tanner graph of the code, and refer to the matched data and ancilla pairs as corresponding qubits.
We then assign a cyclic ordering to the memory blocks and connect each data qubit of each block to the corresponding ancilla qubit of the next block.
We can then implement cyclic shifts of the set of memory blocks using the analogous operation to that used for primitive intra-block automorphisms.
These connections only increase the connectivity degree of each qubit by one, meaning that the connectivity degree of the memory blocks other than the port block is seven, and of the port block is eight.

An alternative construction of the memory could also be used where every memory block is equipped with gadgets to act as a port block.
Bridging these together in a chain then allows for access to any memory window without requiring cyclic shifts.
This approach increases the overhead of the memory blocks by approximately 30\% but, since it removes the need for the connections between blocks described in the previous paragraph, it ensures that no memory block has a connectivity degree greater than seven, and reduces the number of connections that must be added to each memory block to allow memory access from $1020$ to $130$.
Therefore, while both constructions are consistent with the requirement that the connectivity degree is no greater than eight, this alternative construction may be preferred if the benefit of reducing the connectivity in the memory outweighs the increased overhead.

\subsubsection{Memory Sectors}
\label{sec:memory-sectors}
While the above construction is sufficient to incorporate memory, to minimise the overhead we also allow for logical memory qubits to be stored in the second logical sector of the blocks of the processing unit, which we call the \textit{memory sector}.
To allow this, it is sufficient that fan out from any memory sector onto an ancillary processing sector can be performed.
To achieve this, we first define the logical basis of the memory sector to be dual to that of the processing sector, such that logical $Z$ operators correspond to physical $X$ operators.
We then connect a gadget for measuring physical $X$ operators (logical $\bar{Z}$ operators) on the memory sector to a position chosen such that it overlaps with no more than one existing gadget on any physical qubit, and insert this into the bridged chain of processing unit gadgets.
This ensures that it remains the case that no processing block qubit is in the support of more than two gadgets, so that the connectivity degree remains no greater than eight.

By performing a cyclic shift to move the code block into the correct position relative to this gadget, we can then perform an arbitrary $ZZ$-type measurement between any processing sector and memory sector.
In particular, if a processing sector hosting logical ancilla qubits is selected to be the target for the fan out of a memory window, the fan out can then be performed by first using Clifford frame cleaning on the $k/2$ logical qubits of that sector and then using $ZZ$-type measurements to perform the fan out.
A fan in can similarly be performed at the end of the memory access.
This allows the processing unit the required access to any memory window.

\subsection{Magic engine}
\label{sec:magic}
We present a method for preparing and injecting a $|\overline{\text{CCZ}}\rangle$ state with infidelity $2\times 10^{-10}$ in order to perform one Toffoli gate per five logical cycles.
This matches the rate of the standard Pinnacle architecture for factoring, where each Toffoli is compiled from $|\bar{T}\rangle$ states over five logical cycles.

The core of the magic engine is a GB code block with four $Z$-type gadgets and three $X$-type gadgets.
To ensure that the connectivity degree of the code block remains no greater than eight, the gadgets are attached such that no qubit or check is in the support of more than two of them.
However, they are also chosen such that they can measure the set of operators documented in \cref{tab:magic-engine-operators}.
Note that all logical qubits of this GB code refer to the first logical sector; unlike in the baseline architecture, the other logical sector is not used.

\begin{table}
    \centering
    \caption{Logical operators measured by the seven gadgets of the magic engine block under different code block cyclic shifts in each of the five logical cycles of the engine's operation.
    \# labels the number of the logical cycle.
    $\sigma$ denotes the size of the cyclic shift of the code block for that logical cycle relative to the initial configuration.
    $G_{Z,i}$ for $i=1,2,3,4$ denote the four $Z$-type gadgets and $G_{X,i}$ for $i=1,2,3$ denotes the three $X$-type gadgets.
    All operators are on the first logical sector.
    Dashes indicate that that gadget is not activated for that logical cycle.}
    \label{tab:magic-engine-operators}

    \setlength{\tabcolsep}{2pt}
    \renewcommand{\arraystretch}{1.1}

    \resizebox{\columnwidth}{!}{%
    \begin{tabular}{|c|c|c|c|c|c|c|c|c|}
    \hline
       \# & $\sigma$
       & $G_{Z,1}$ & $G_{Z,2}$ & $G_{Z,3}$ & $G_{Z,4}$
       & $G_{X,1}$ & $G_{X,2}$ & $G_{X,3}$ \\
    \hline
    1 & 0
      & $\bar{Z}_2\bar{Z}_4$
      & $\bar{Z}_1\bar{Z}_2\bar{Z}_3\bar{Z}_4$
      & $\bar{Z}_1\bar{Z}_4$
      & $\bar{Z}_3\bar{Z}_4$
      & - & - & - \\
    \hline
    2 & 85
      & $\bar{Z}_1\bar{Z}_2\bar{Z}_4$
      & $\bar{Z}_4$
      & $\bar{Z}_{1}\bar{Z}_3\bar{Z}_4$
      & $\bar{Z}_2\bar{Z}_3\bar{Z}_4$
      & - & - & - \\
    \hline
    3 & 30
      & - & - & - & -
      & - & - & $\bar{X}_4$ \\
    \hline
    4 & 170
      & $\bar{Z}_1$
      & -
      & $\bar{Z}_3$
      & $\bar{Z}_2$
      & - & - & - \\
    \hline
    5 & 170
      & - & - & - & -
      & $\bar{X}_1$
      & $\bar{X}_2$
      & $\bar{X}_3$ \\
    \hline
    \end{tabular}%
    }
\end{table}

Since the operation of the engine is periodic with a period of five logical cycles, we enumerate the logical cycles of a period 1--5.
During logical cycles 1, 2 and 3 we perform 8T-to-CCZ magic state distillation using the circuit presented in Figure 16b of Ref.~\cite{litinski_magic_2019}.
Logical cycles 1 and 2 each consist of parallel auto-corrected injection of four noisy $|T\rangle$ states from distance-15 surface codes.
These injections are done through parallel measurements using each of the four $Z$-type GB code gadgets joint to $Z$ surface code measurements, along with joint measurement between each surface code and an ancillary surface code used for auto-correction \cite{litinski_magic_2019}.
The noisy $|T\rangle$ states themselves can be prepared with an infidelity of $2.7\times 10^{-6}$ using magic state cultivation \cite{gidney_magic_2024}, with a sufficiently low discard rate to allow an overall engine reject rate of 1\%; this process is detailled in appendix \ref{sec:appendix-b}.
During logical cycle 3, one $X$-type gadget is used to measure $\bar{X}_4$ to allow for post-selection and complete the distillation circuit.
The result is that the first three logical qubits are prepared in the $|\overline{\text{CCZ}}\rangle$ state
with infidelity $p_{\text{out}}\approx28p_{\text{in}}^2=2\times 10^{-10}$ with a $p_r=1\%$ reject rate.

In logical cycles 4 and 5, the state is teleported onto three ancillary logical qubits of the processing unit, which we denote by $1,a$, $2,a$ and $3,a$ respectively.
These three ancillary logical qubits can each initially be assumed to be in the $|\bar{0}\rangle$ state (up to a Pauli frame correction since, as stated below, they are each measured in the $\bar{Z}$ basis in logical cycle 3).
They are chosen such that each lies on a different processing block, and they used exclusively for the purpose described here to ensure that their Clifford frame remains trivial.
An additional $X$-type gadget is added to each of these processing blocks at a position chosen such it remains the case that no qubit or check connects to more than two gadgets, which ensures that the fixed connectivity degree of each block remains no greater than eight.
We refer to these three gadgets collectively as a \textit{magic port}.
Each of the magic port gadgets is bridged to one of the $Z$-type magic engine gadgets (specifically, $G_{Z,1}$, $G_{Z,4}$ and $G_{Z,3}$ with reference to \cref{tab:magic-engine-operators}); they are not bridged to any other gadgets, since they are exclusively used for this state teleportation.
This setup allows parallel measurement of  $\bar{Z}_1\bar{X}_{1,a}$, $\bar{Z}_2\bar{X}_{2,a}$ and $\bar{Z}_3\bar{X}_{3,a}$, which is performed in logical cycle 4.
In logical cycle 5, we then measure $\bar{X}_1$, $\bar{X}_2$ and $\bar{X}_3$, which inform Pauli frame updates on the ancillary logical qubits.
This amounts to teleportation up to a logical Hadamard, such that the three ancillary logical qubits are prepared in the state $\bar{H}^{\otimes 3}|\overline{\text{CCZ}}\rangle$.

The $\bar{H}^{\otimes 3}|\overline{\text{CCZ}}\rangle$ state is then injected from the three ancillary logical qubits to perform an arbitrary Toffoli on the processing unit.
This is done in logical cycles 5, 1, 2 and 3 (i.e., those following the joint $ZX$-type teleportation measurements).
Specifically, let $\bar{C}$ be the Clifford frame on the processing unit, and label the three logical qubits on which the Toffoli acts as $c_1$, $c_2$ and $t$ respectively.
Then, the required injection can be performed (using a circuit based on that from Figure 14a of Ref.~\cite{litinski_active_2022}) by measuring
$\bar{X}_{1,a}\cdot\bar{C}\bar{Z}_{c_1}\bar{C}^\dag$, $\bar{X}_{2,a}\cdot\bar{C}\bar{Z}_{c_2}\bar{C}^\dag$, $\bar{X}_{3,a}\cdot\bar{C}\bar{X}_t\bar{C}^\dag$ in series across the logical cycles 5,1 and 2 respectively, and then measuring $\bar{Z}_{1,a}$, $\bar{Z}_{2,a}$ and $\bar{Z}_{3,a}$ in parallel in the logical cycle 3 (and absorbing appropriate Clifford corrections into the Clifford frame of the processing unit).
These injection measurements are all performed using the existing gadgets of the processing unit (not the magic port); the parallel measurements on separate blocks in logical cycle 3 are enabled by not activating the bridges between blocks in that logical cycle.

This construction maintains the property that each gadget is connected at a fixed position and bridged to at most two other gadgets, ensuring that the fixed connectivity degree remains eight.

\section{Simulation Results}
\label{sec:simulations}
\begin{figure}
  \centering
  \includegraphics[width=\linewidth]{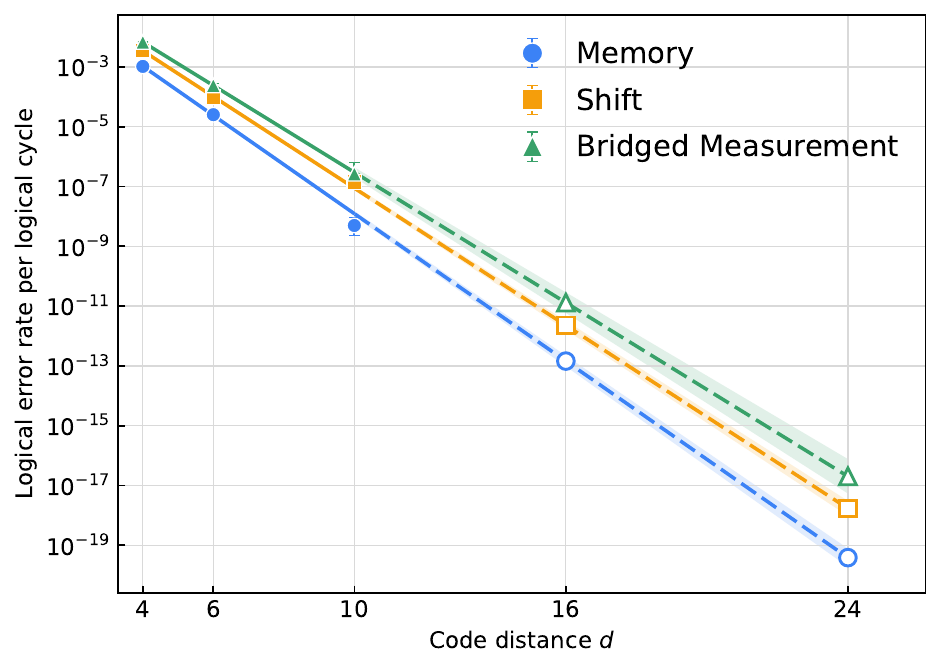}
  \caption{Circuit-level simulation results for the per-block logical error rate of the Frozen Pinnacle components at \(p=10^{-3}\), versus code distance \(d\).
    Filled markers are simulated data (\(d=4,6,10\)) while open markers (\(d=16,24\)) are extrapolated from the fit of \cref{eq:ansatz}.
    Error bars and the shaded extrapolation band denote \(95\%\) confidence intervals.}
  \label{fig:simulations}
\end{figure}

We assess the performance of the Frozen Pinnacle Architecture through noisy circuit-level simulations using standard circuit-level depolarising noise at a physical error rate of $p=10^{-3}$.
Due to the infeasibility of sampling at the low error rates achieved by the full-distance architecture, we perform these simulations for the corresponding architectures that use the $[[30,8,4]]$, $[[62,10,6]]$ and $[[126,12,10]]$ GB codes from Ref.~\cite{webster_pinnacle_2026} and extrapolate from these results.

We simulate the operations necessary for arbitrary logical Pauli measurements on processing sectors with fixed connectivity degree eight, namely shift automorphisms of a code block and logical measurement using $k/2+1$ bridged gadgets.
This is motivated by the fact that this capability is the essential foundation on which the Frozen Pinnacle Architecture depends.
We compare these operations to a memory experiment which serves as a baseline for the performance of the code block in the absence of these operations.

The results of these simulations are shown in \cref{fig:simulations}.
The series of this figure represent the following:
\begin{itemize}
\item \textit{Memory}: memory experiment on the generalised bicycle code block via repeated rounds of noisy syndrome extraction of the code stabiliser generators.
\item \textit{Shift}: shift automorphism circuit during repeated rounds of noisy syndrome extraction.
  For each code, the maximum number of primitive shifts $w$ required to measure an arbitrary logical Pauli operator on the $k/2$ qubits of the processing sector is constructed.
\item \textit{Bridged Measurement}: joint measurement of the logical operators measured by the single $X$-type gadget and the $k/2$ $Z$-type gadgets via generalised surgery.
Since measuring an arbitrary logical Pauli operator on the processing sector requires activating a subset of these gadgets, this serves as a conservative estimate of the performance of such a measurement.
\end{itemize}

In each experiment, we simulated $d_t=d+2$ rounds of syndrome extraction.
Since the memory and bridged measurement simulations contain no shift automorphisms, their logical error rates are rescaled to the full logical-cycle length of $d_t+w$, accounting for the additional code cycles required when the shift automorphisms are applied.
The shift automorphism circuit simulations are rescaled to $d_t+w-1$ to account for the extra rounds of syndrome extraction between each shift.

The syndrome extraction circuit for the memory is scheduled as in Ref.~\cite{lin_singleshot_2025}.
The shift automorphism circuits are performed via two layers of SWAP gates between ancilla and data qubits implemented using two CNOTs each.
The ancillae are reset and measured in their respective basis before and after each of these shifts and the measurements included as detectors for error detection and decoding.
The syndrome extraction for the bridged surgery experiment is constructed by fixing the bulk schedule of the GB code and scheduling the remaining checks using a depth-optimal CP-SAT formulation.

To obtain estimates for the $[[254,14,16]]$ and $[[510,16,24]]$ GB codes, we assume each point is binomially distributed and fit the results to the sub-threshold ansatz
\begin{equation}
  \label{eq:ansatz}
p^{\star}_{L,cb}(d) = A\left(\frac{10^{-3}}{B}\right)^{\frac{d}{2}},  
\end{equation}

where $p^{\star}_{L,cb}(d)$ is the logical failure rate of any of the $k$ logical observables over the $d_t+w$ rounds at the fixed physical error rate $p=10^{-3}$.
The fitted parameters for the three simulated experiments are given in \cref{tab:fit-params}.
The extrapolated points in \cref{fig:simulations} are obtained from that ansatz.
Since an arbitrary logical measurement requires both a shift automorphism and a bridged measurement, we conservatively sum their contributions, giving an estimated error rate \emph{per logical qubit} per logical cycle of $1.4\times 10^{-18}$ ($95\%$ confidence interval $[4.3\times 10^{-19}, 4.9\times 10^{-18}]$) for the $d=24$ code.
This is significantly below the logical error rate of approximately $10^{-14}$ per logical qubit per logical cycle required for RSA-2048 factoring \cite{webster_pinnacle_2026}, justifying that this application is supported on the architecture.

\begin{table}
  \centering
  \caption{Fitted parameters of the ansatz in \cref{eq:ansatz} with 95\% confidence intervals for the simulated components.}
  \label{tab:fit-params}
  \begin{tabular*}{\columnwidth}{@{\extracolsep{\fill}} l c c @{}}
    \toprule
    Experiment & $A$ & $B$ \\
    \midrule
    Memory & $2.05^{+0.37}_{-0.32}$ & $0.0440^{+0.0033}_{-0.0031}$ \\
    Shift  & $4.01^{+0.54}_{-0.48}$ & $0.0340^{+0.0020}_{-0.0019}$ \\
    Bridged Measurement & $5.49^{+1.92}_{-1.42}$ & $0.0284^{+0.0041}_{-0.0036}$ \\
    \bottomrule
  \end{tabular*}
\end{table}

\section{RSA-2048 Factoring Resource Estimate}
\label{sec:resource-estimate}
We now perform a resource estimate for factoring on the Frozen Pinnacle Architecture.
We follow the compilation of Ref.~\cite{webster_pinnacle_2026} (which itself follows Ref.~\cite{gidney_how_2025}), but allow only a single processing unit.

We denote the number of working register logical qubits required for factoring in the baseline architecture, which is given by equation 23 of Ref.~\cite{webster_pinnacle_2026}, by $\kappa_b$.
The number of logical qubits required for the modified architecture presented here is then $\kappa=\kappa_b+2$, which accounts for the three ancillary logical qubits used for $|\overline{\text{CCZ}}\rangle$ injection, which replace one ancilla qubit used to synthesise Toffolis from $|\bar{T}\rangle$ states in the baseline architecture.
Additionally, $m$ input register logical qubits are required as specified in Ref.~\cite{gidney_how_2025,webster_pinnacle_2026}.
When the factoring algorithm is compiled to this architecture, each processing block contributes $k/2=8$ working register logical qubits (i.e., those in the processing sector) and $k/2=8$ input logical qubits (i.e., those in the memory sector), and each memory block contributes $k=16$ input logical qubits.
This means that $\lceil \kappa/8 \rceil$ processing blocks and $\lceil m/16-\lceil\kappa/8\rceil/2 \rceil=\lceil \left(m-8\lceil \kappa/8\rceil\right)/16\rceil$ memory blocks provide the required logical qubits.
In addition, we require one memory port, one magic engine (with a magic port).

Letting $n_{pb}$, $n_{mb}$, $n_{me}$ and $n_p$ be the number of physical qubits per processing block, memory block, magic engine and memory port respectively, it therefore follows that the number of physical qubits is given by:
\begin{equation}
n=\lceil \kappa/8\rceil n_{pb}+\lceil \left(m-8\lceil \kappa/8\rceil\right)/16\rceil n_{mb}+n_p+n_{me}.
\end{equation}

A processing block consists of a GB code block ($1020$ qubits) and $k/2+2=10$ gadgets, of which nine have 99 qubits and one has 101 qubits, with bridges ($9\cdot99+101+10\cdot 51)=1502$ qubits), so $n_{pb}=2522$.  

A memory block consists of a GB code block, so $n_{mb}=1020$ and a port consists of two gadgets (each with 99 qubits) with a bridge between them, so $n_p=2\cdot 99+51=249$. Note that the bridge between the second port gadget and the processing unit is already accounted for in the processing blocks, since it is the bridge associated with the last gadget of the processing unit.

A magic engine consists of a GB code block with four $Z$-type gadgets which each have 99 qubits and three $X$-type gadgets which each have 101 qubits, totalling $1020+4\cdot 99+3\cdot 101=1719$.
It additionally consists of twelve $454$ qubit ancillary surface code patches and four pairs of bridges of $2\cdot 15-1=29$ qubits each to allow auto-corrected injection into the magic engine, adding a further $12\cdot 454+8\cdot 29=5680$ qubits.
Finally, it also requires a magic port consisting of three $X$-type gadgets which each have $101$ qubits, with bridges which each have $51$ qubits, adding a further $3(101+51)=456$.
The magic engine therefore requires a total of $n_{me}=1719+5680+456=7855$ qubits.

The total number of physical qubits is therefore:
\begin{equation}
n=2522\lceil\kappa/8\rceil +1020\lceil \left(m-8\lceil \kappa/8\rceil\right)/16\rceil+8104.
\end{equation}

The number of logical cycles required is the same as in the baseline architecture, aside from adjustments that account for 1) that only one $|\overline{\text{CCZ}}\rangle$ magic state is required per Toffoli instead of four $|\bar{T}\rangle$ states but that these states have a higher infidelity ($p_{\text{out}}=2\times 10^{-10}$), 2) the different reject rate of the magic engines ($p_r=0.01$), and 3) that we assume 5/6 of logical cycles can be delayed by magic engine (i.e., those which are part of implementing a Toffoli) instead of 4/6 in the baseline architecture.
All of these adjustments are included in our estimate, but their effects are marginal.
In addition, as noted in \cref{sec:processing-unit}, the requirement of cyclic shift automorphisms increases the number of code cycles per logical cycle from $d_t=d+2=26$ in the baseline architecture to $d_t+4=d+6=30$.

We minimise the number of physical qubits over the free parameters of the algorithm, under the constraint that the runtime is no greater than one month, with a code cycle time of 1 \textmu s and a reaction time of 10 \textmu s.
This gives $\kappa=144$ and $m=1195$, resulting in a total of $n=120\,820$ physical qubits.

\section{Conclusion}
We have shown that the Pinnacle Architecture can be adapted to allow for a 2048-bit integer to be factored in one month with approximately 120\,000 physical qubits with a fixed connectivity degree of eight.
This result demonstrates that the constraint of fixed connectivity does not intrinsically necessitate a significant increase in time overhead of the kind incurred by the bicycle architecture of Ref.~\cite{yoder_tour_2025}.
This affirms that the key advantage of the Pinnacle Architecture is its ability to measure any logical Pauli operator in a single logical cycle, which avoids the substantial compilation overhead incurred from use of a more restricted instruction set.
More generally, it supports the conclusion that the low spacetime overhead observed in Ref.~\cite{webster_pinnacle_2026} is not substantially an artefact of its hardware-agnostic assumptions.

\section*{Acknowledgements}
We thank all our colleagues at Iceberg Quantum for helpful discussions.

\appendix

\section{Magic State Cultivation}
\label{sec:appendix-b}
In preparing the noisy magic states for input into the magic engine in \cref{sec:magic}, we use the magic state cultivation protocol of Ref.~\cite{gidney_magic_2024}.
The data published with that work shows that $|\bar{T}\rangle$ states with $2.7\times 10^{-6}$ infidelity can be produced using the scheme with fault-distance three with a success probability of 0.23.
An expected spacetime volume of approximately 8000 qubit-rounds per successful $|\bar{T}\rangle$ state preparation is shown in Figure 1 of that work.
Taking into account that there are four logical cycles of $30$ GB code cycles (which we conservatively equate with cultivation rounds) each between injections of each state, a naive estimate would therefore suggest that approximately $(8\cdot 8000)/(4\cdot 30)=533$ physical qubits would suffice to provide the required throughput.
However, we note that this is an optimistic estimate, since it does not account for latency resulting from variance in the true number of qubit-rounds required in each attempt, or for packing constraints.

To address this, we instead base our results on the following  conservative analysis.
First, we use twelve separate ancilla patches which each contain 454 physical qubits -- enough to host the $d=15$ grafted surface code that results in successful cultivations.
On each patch, we perform four parallel attempts of the first two stages (injection and cultivation) on sets of 15 qubits at the corner of each patch; these positions are chosen to ensure compatibility of any of the output states with the subsequent escape stage, without requiring any additional routing.
The reject rate prior to escape for each of the four attempts is 0.35, meaning that each patch produces at least one state that is accepted for escape with very high probability.
Assuming there exists such an accepted state, it can then be escaped to the $d=15$ surface code and the full cultivation protocol completed, which -- conditional on no discard before escape -- has a success rate of 0.23/0.65=0.35.
Any patch that successfully prepares an accepted state is then left idle storing this state until injection, while the other patches are reinitialised for a new attempt.
Following Figure 15 of Ref.~\cite{gidney_magic_2024}, each end-to-end attempt requires 18 rounds, meaning that up to five such attempts can be completed within three 30-round GB logical cycles.
Each attempt has a success probability of $0.35$, which means that each patch has a probability of $1-0.65^5=0.88$ of successfully preparing a state in this time, and hence the probability that at least eight of the twelve patches successfully prepare a state is $\sum_{s=8}^{12}\binom{12}{s}(0.88)^s(0.12)^{12-s}=0.99$.
If this does not occur, then we conservatively assume the entire distillation to have failed; since the post-selection reject rate of the distillation circuit itself (approximately $8p_{\text{in}}=0.002\%$) is negligible compared with this, this sets the overall engine reject rate of $p_r=1\%$.  
Otherwise, we use eight patches with successfully prepared states for injection and use the other four as ancillae for auto-correction (which are reused across the two batches of injections).
We assume that the routing required to ensure that $|\bar{T}\rangle$ states are in the required patches is negligible, which is justified since we can allow for transversal swaps between nearest-neighbour patches in a line while retaining a connectivity degree of below eight.

\bibliography{refs}

\end{document}